\documentclass[aps,prl,twocolumn,groupedaddress,showpacs]{revtex4}
\usepackage{bm}
\usepackage{graphicx}
\usepackage{amsmath}
\usepackage{color}
\def\Vec#1{\mbox{\boldmath $#1$}} 	
\begin{document}

\title{Tunability of the $\Vec{k}$-space Location of the Dirac Cones in the Topological Crystalline Insulator Pb$_{1-x}$Sn$_x$Te}
\author{Y. Tanaka,$^1$ T. Sato,$^1$ K. Nakayama,$^1$ S. Souma,$^2$ T. Takahashi,$^{1,2}$ Zhi Ren,$^3$ M. Novak,$^3$ Kouji Segawa,$^3$ and Yoichi Ando$^3$}
\affiliation{$^1$Department of Physics, Tohoku University, Sendai 980-8578, Japan}
\affiliation{$^2$WPI Research Center, Advanced Institute for Materials Research,
Tohoku University, Sendai 980-8577, Japan
}
\affiliation{$^3$Institute of Scientific and Industrial Research, Osaka University, Ibaraki, Osaka 567-0047, Japan}

\date{\today}

\begin{abstract}
We have performed systematic angle-resolved photoemission spectroscopy of the topological crystalline insulator (TCI) Pb$_{1-x}$Sn$_x$Te to elucidate the evolution of its electronic states across the topological phase transition.  As previously reported, the band structure of SnTe ($x = 1.0$) measured on the (001) surface possesses a pair of Dirac-cone surface states located symmetrically across the $\bar{X}$ point in the (110) mirror plane. Upon approaching the topological phase transition into the trivial phase at $x_c\approx0.25$, we discovered that Dirac cones gradually move toward the $\bar{X}$ point with its spectral weight gradually reduced with decreasing $x$. In samples with $x \leq 0.2$, the Dirac-cone surface state is completely gone, confirming the occurrence of the topological phase transition. Also, the evolution of the valence band feature is found to be consistent with the bulk band inversion taking place at $x_c$. The tunability of the location of the Dirac cones in the Brillouin zone would be useful for applications requiring Fermi-surface matching with other materials, such as spin injection. 
\end{abstract}

\maketitle
  
      Topological insulators (TIs) materialize a new quantum state of matter where an unusual metallic state protected by time-reversal symmetry (TRS) appears at the edge or the surface of a band insulator \cite{HasanReview, SCZhangReview}. The discovery of TIs triggered the search for new types of topological materials protected by other symmetries \cite{SchnyderPRB, Kitaev, RangArXiv, MongPRB, WangNP}, and a recent theory predicted the existence of ``topological crystalline insulators'' (TCIs) in which metallic surface states are protected by point-group symmetry of the crystal structure \cite{FuTCIPRL}; in particular, mirror symmetry leads to the topological invariant called mirror Chern number \cite{FuTCINC}. Such a TCI phase has been experimentally verified by angle-resolved photoemission spectroscopy (ARPES) experiments for narrow-gap IV-VI semiconductors SnTe \cite{TanakaNP}, Pb$_{0.6}$Sn$_{0.4}$Te \cite{HasanTCINC}, and Pb$_{0.77}$Sn$_{0.23}$Se \cite{DziawaNM}, all having the same crystal structure shown in Fig. 1(a). In those materials, the topological surface states measured on the (001) surface consist of Dirac cones located at momenta slightly away from the time-reversal-invariant momentum (TRIM) $\bar{X}$ point in the (110) mirror plane of the crystal [Fig. 1(b)]; since there are four $\bar{X}$ points on the boundary of the first surface Brillouin zone (BZ), there are a total of four Dirac cones within it \cite{TanakaNP, HasanTCINC, DziawaNM}. This is distinct from the three-dimensional (3D) TIs whose surface states are characterized by an odd number of Dirac cones \cite{Z2FuKane}, although the Dirac-cone surface states in TIs and TCIs are both produced by bulk band inversion caused by a strong spin-orbit coupling. Note that, due to the periodicity of the BZ, one always sees a pair of Dirac cones across $\bar{X}$ in TCIs, which leads to a characteristic ``double Dirac-cone'' structure involving a Lifshitz transition at high binding energies \cite{FuTCIPRL,TanakaNP}.
     
   In contrast to the double Dirac-cone signature observed in the TCI phase, the ARPES measurements for isostructual PbTe \cite{TanakaNP} and Pb$_{0.8}$Sn$_{0.2}$Te \cite{HasanTCINC} revealed the absence of any surface states, which strongly suggests a trivial-to-nontrivial topological quantum phase transition (QPT) in the solid-solution system Pb$_{1-x}$Sn$_x$Te \cite{TanakaNP,HasanTCINC}. This is in line with the established band diagram of this compound which shows a band inversion somewhere near $x \sim 0.3$ \cite{DimmockPRL,AkimovPSS} that is supported by band-structure calculations \cite{FuTCINC,GaoPRB}. However, it is still unclear how exactly the surface and bulk electronic states evolve as a function of Sn composition $x$, which would be useful for establishing practical understanding of TCIs. In fact, before Pb$_{1-x}$Sn$_x$Te became relevant in the topological context, there were no ARPES studies of the evolution of its electronic structure in the mixed compositions despite the continued interest in this system in the semiconductor community \cite{AkimovPSS,GaoPRB}.
   
   In this Letter, we report comprehensive ARPES studies of Pb$_{1-x}$Sn$_x$Te with various Sn compositions ($x$ = 0.0, 0.2, 0.3, 0.5, 0.6, 0.8, and 1.0). The double Dirac-cone surface state persists in the whole TCI phase of $x \geq 0.3$, but surprisingly, the $\Vec{k}$ position of the Dirac point (called Dirac momentum) was found to systematically move within the mirror plane toward the $\bar{X}$ point with decreasing $x$. This means that one can $tune$ the Dirac momentum, albeit in a limited range, by changing the composition in this system, which is new to a Dirac system. In addition, we were able to trace the bulk-band dispersion in the 3D $\Vec{k}$ space, and found that the spectral signature is consistent with the band inversion at the QPT.

\begin{figure}
\includegraphics[width=3.4in]{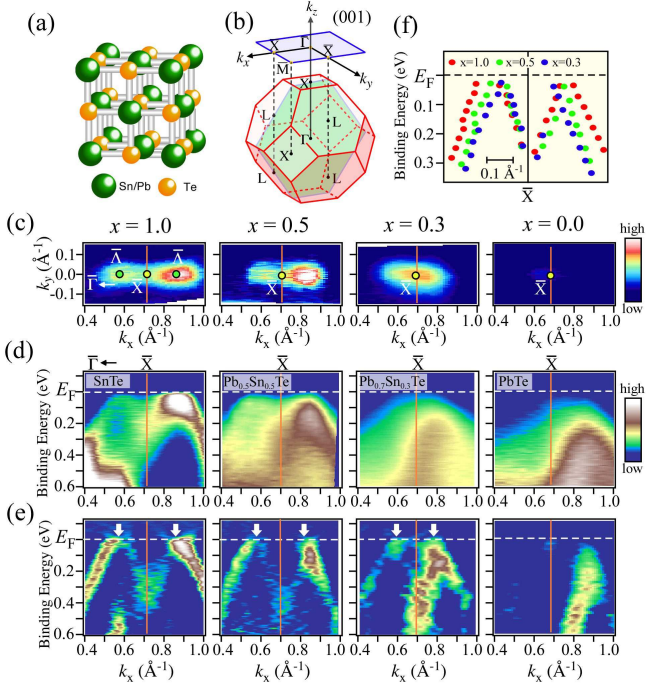}
\vspace{-0.4cm}
\caption{(Color online): (a) Crystal structure of Pb$_{1-x}$Sn$_x$Te. (b) Bulk fcc BZ and corresponding tetragonal (001) surface BZ. The (110) mirror plane is indicated by the green shaded area. (c) Mapping of ARPES intensity at $E_{\rm F}$ around the $\bar{X}$ point for $x$ = 1.0, 0.5, 0.3, and 0.0 plotted as a function of in-plane wave vector ($k_x$ and $k_y$) measured with the He I$\alpha$ line ($h\nu$ = 21.218 eV) at $T$ = 30 K. The intensity is obtained by integrating the spectra within $\pm$10 meV of $E_{\rm F}$. (d) Corresponding near-$E_{\rm F}$ ARPES intensity along the $\bar{\Gamma}\bar{X}$ cut. (e) Band dispersions derived from the second derivatives of the momentum distribution curves (MDCs) along the $\bar{\Gamma}\bar{X}$ cut. The $\Vec{k}$ location of the Dirac point (the $\bar{\Lambda}$ point) is indicated by white arrows in (e). (f) Comparison of the band dispersions extracted from the peak position of the second derivatives of the MDCs for $x$ = 1.0, 0.5, and 0.3.}
\end{figure}
   
    High-quality single crystals of the two end-member samples SnTe ($x$ = 1.0) and PbTe ($x$ = 0.0) were grown by the modified Bridgeman method \cite{TanakaNP}, and the solid-solution samples ($x$ = 0.2, 0.3, 0.5, 0.6, and 0.8) were prepared by a vapor transport method. High-purity elements of Sn (99.99{\%}), Pb (99.998{\%}), and Te (99.999{\%}) are used for the growths. The actual Sn content $x$ in the grown crystals was measured by the inductively-coupled plasma atomic emission spectroscopy and was found to be always consistent with the nominal $x$ (which is shown in this paper) within $\pm$0.02. All the Pb$_{1-x}$Sn$_x$Te single crystals are single domain and show good crystallinity in x-ray Laue analyses. ARPES measurements were performed with the MBS-A1 and VG-Scienta SES2002 electron analyzers with a high-intensity helium (He) discharge lamp at Tohoku University and also with tunable synchrotron lights at the beamline BL28A at the Photon Factory (KEK). To excite photoelectrons, we used the He I$\alpha$ resonance line ($h\nu$ = 21.218 eV) and the circularly polarized lights of 75-100 eV at Tohoku University and the Photon Factory, respectively. The energy and angular resolutions were set at 10-30 meV and 0.2$^{\circ}$, respectively. Samples were cleaved {\it in situ} along the (001) crystal plane in an ultrahigh vacuum of 1$\times$10$^{-10}$ Torr. All the measured samples present shiny mirror-like surface after cleaving, indicative of their high quality. The Fermi level ($E_{\rm F}$) of the samples was referenced to that of a gold film evaporated onto the sample holder.
    
\begin{figure}
\includegraphics[width=3.4in]{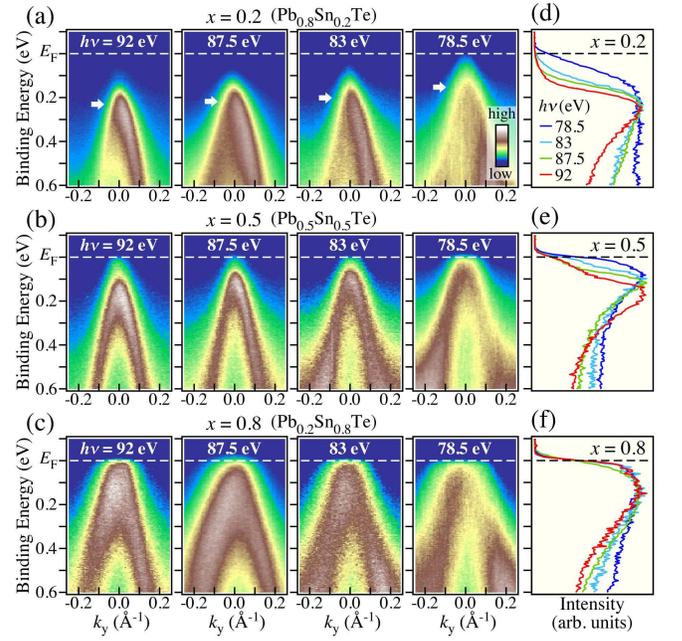}
\vspace{-0.4cm}
\caption{(Color online): (a)-(c) Photon-energy dependence of near-$E_{\rm F}$ ARPES intensity map around the $\bar{X}$ point along a cut across the $\Vec{k}$ point where the bulk VB is located closest to $E_{\rm F}$ at each $h\nu$, shown for $x$ = 0.2, 0.5, and 0.8, respectively. (d)-(f) Comparison of the energy distribution curves (EDCs) at $k_y$ = 0 extracted from the intensity plots for $x$ = 0.2, 0.5, and 0.8, respectively.}
\end{figure}
    
\begin{figure*}
\includegraphics[width=6.4in]{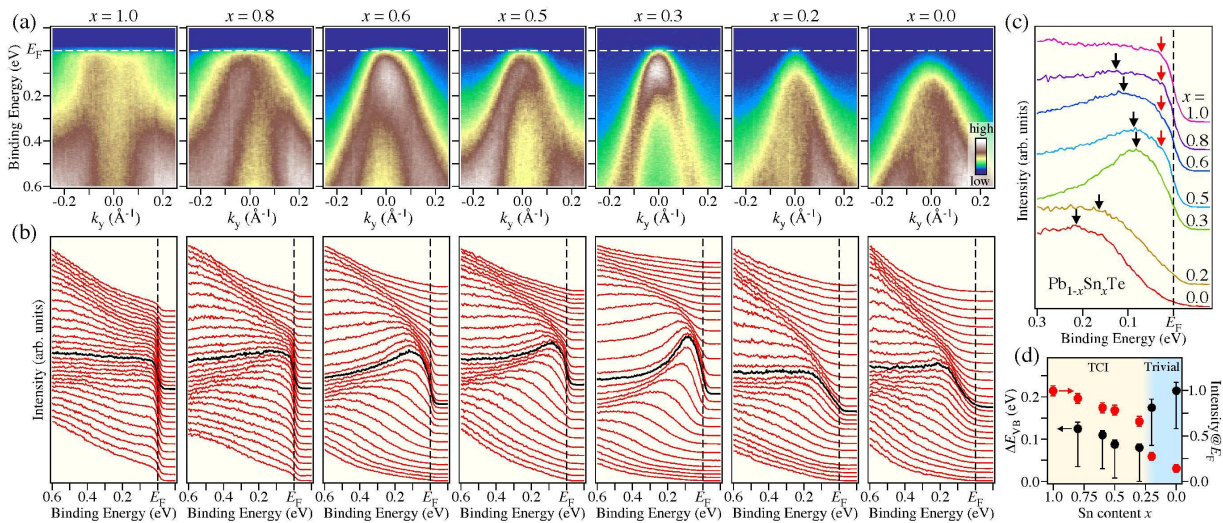}
\vspace{-0.4cm}
\caption{(Color online): (a), (b) $x$ dependence of near-$E_{\rm F}$ ARPES intensity and corresponding EDCs, respectively, measured along the $k_y$ cut across the bulk VB top that can be accessed with $h\nu$ = 78.5 eV. (c) Comparison of the EDCs at $k_y$ = 0 [thick curves in (b)]. Black and red arrows denote the features stemming from the bulk VB and the surface state, respectively. (d) Plots of the $\Delta{E}_{\rm VB}$ value (black circles) and the intensity at $E_{\rm F}$ (red circles) as a function of $x$. The intensity at $E_{\rm F}$ is normalized to that at the VB top.}
\end{figure*}

   Figure 1(c) displays the Fermi-surface (FS) mapping around the $\bar{X}$ point of the surface BZ measured with the He I$\alpha$ resonance line ($h\nu$ = 21.218 eV) for Pb$_{1-x}$Sn$_x$Te with four representative Sn compositions $x$ including both end members SnTe ($x$ = 1.0) and PbTe ($x$ = 0.0). At $x$ = 1.0, one can immediately notice a dumbbell-shaped intensity pattern elongated along the $\bar{\Gamma}\bar{X}$ direction ($k_x$ direction) with its intensity maxima located away from the $\bar{X}$ point. This feature arises from the double Dirac-cone surface state \cite{TanakaNP} whose Dirac points are located at each intensity maxima (defined the $\bar{\Lambda}$ point), as one can see from the near-$E_{\rm F}$ band dispersion along the $\bar{\Gamma}\bar{X}$ cut in Fig. 1(d) showing the band maxima on both sides of the $\bar{X}$ point. Such an ``M''-shaped band dispersion is characteristic of the surface state since it is absent in the bulk-band calculations \cite{FuTCINC, HasanTCINC, TungPR, MelvinJPhysC, LittlewoodPRL}. The M-shaped dispersion and the dumbbell-shaped FS are also observed for $x$ = 0.5, which signifies that the system still belongs to the TCI phase, although the intensity distribution around $E_{\rm F}$ appears to be weaker and broader than that for $x$ = 1.0. At $x$ = 0.3, however, the observed ARPES intensity looks different, and neither the dumbbell-shaped FS nor the M-shaped dispersion are clearly resolved. This is because the relative intensity of the bulk band gains strength and the surface band is becoming obscured. Nevertheless, we are still able to trace the surface-band dispersion by taking second derivatives of the intensity along the $k_x$ direction and, as shown in Fig. 1(e), the M-shaped dispersion is still discernible. This suggests that $x$ = 0.3 is in the vicinity of the QPT but is still on the TCI side. In contrast, the spectral feature for $x$ = 0.0 looks substantially different. The near-$E_{\rm F}$ intensity is strongly suppressed and no Dirac-cone surface band is observed, reflecting the topologically trivial (ordinary) nature of PbTe \cite{TanakaNP}.

The most important feature found in Fig. 1 is the evolution of the surface state in the TCI phase. In particular, as one can see in Fig. 1(e), the $\Vec{k}$ position of the Dirac point in the BZ ($i.e.$, the $\bar{\Lambda}$ point which we also call Dirac momentum) systematically $moves$ toward the $\bar{X}$ point with decreasing $x$ (see white arrows). A comparison of the extracted band dispersions shown in Fig. 1(f) further reveals that the slope of the Dirac cones (Fermi velocity) is essentially unchanged despite the systematic shift of the Dirac momentum. This peculiar evolution of the double Dirac-cone structure in the TCI phase is the main finding of this work.

   Since a topological phase transition must be accompanied by a band gap closing, it is useful to look for its signature in the ARPES data. We have performed photon-energy dependent ARPES measurements for each $x$ and determined the bulk-band dispersion along the $k_z$ (wave vector perpendicular to the surface) direction. The representative band dispersions obtained near the $\bar{\Lambda}$ point along the $k_y$ direction (perpendicular to the $\bar{\Gamma}\bar{X}$ direction) for selected $x$ values (0.2, 0.5, and 0.8) are shown in Figs. 2(a)-(c). At $x$ = 0.2, there exists a highly dispersive band showing a holelike dispersion whose top is dependent on the photon energy; it is located at the binding energy ($E_{\rm B}$) of 0.22 eV at $h\nu$ = 92 eV, but it gradually moves toward $E_{\rm F}$ (see white arrows) with decreasing $h\nu$. This trend can also be seen in Fig. 2(d) where the energy distribution curves (EDCs) for $k_y$ = 0 measured with various photon energies from $h\nu$ = 78.5 to 92 eV are overlaid, which clearly presents a peak shift. We found a similar trend for $x$ = 0.5 [Fig. 2(b)], although the energy location of the band is quantitatively different. Similar to the case of $x$ = 0.2, the prominent band at $x$ = 0.5 is attributed to the bulk valence band (VB) because it shows a finite $k_z$ dispersion [see also Fig. 2(e) for the peak shift in EDCs]. On the other hand, the $k_z$ dispersion is not so obvious for $x$ = 0.8 [Figs. 2(c) and (f)], likely due to the stronger surface-band intensity and relative suppression of the bulk-band weight, as we discuss in more detail later.
   
   We have previously elucidated that $h\nu\sim78.5$ eV hits to the right $k_z$ value to probe the L point of the bulk BZ projected to the $\bar{X}$ point of the surface BZ and, hence, is maximally suited to measure the top of the bulk VB in the 3D $\Vec{k}$ space \cite{TanakaNP} (note that the bulk FS is expected to be located at the L point in this system \cite{MelvinJPhysC, LittlewoodPRL, SatoInSnTe}). Therefore, to elucidate the evolution of the bulk VB as a function of $x$, we have measured all the samples with $h\nu$ = 78.5 eV along the $\Vec{k}$ cut to cross the bulk VB top. Figures 3(a) and (b) show the resulting near-$E_{\rm F}$ ARPES intensity and corresponding EDCs, respectively. We note that the features originating from deeper bands are observed at nearly the same binding energy in all the samples, suggesting that we can compare the positions of the VB top without taking into account the $x$ dependence of the chemical potential within the experimental uncertainty \cite{note}.
   
       \begin{figure}
\includegraphics[width=3.2in]{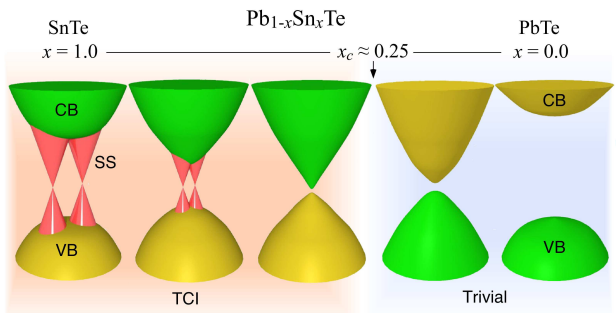}
\vspace{-0.4cm}
\caption{(Color online): Schematic picture of the energy band evolution around the $\bar{X}$ point in Pb$_{1-x}$Sn$_x$Te summarizing the present ARPES experiments. CB, VB, and SS denote the bulk conduction band, bulk valence band, and surface state, respectively. }
\end{figure}
   
At $x$ = 1.0, we observe a highly dispersive holelike band which is attributed to an admixture of bulk and surface bands with dominant contribution from the surface state near $E_{\rm F}$ \cite{TanakaNP, SatoInSnTe}. In fact, the EDCs for $x$ = 1.0 exhibit a clear Fermi-edge cutoff which is a fingerprint of the metallic topological surface state. The Fermi edge can be clearly recognized down to $x$ = 0.5 in the EDCs at $k_y$ = 0 summarized in Fig. 3(c) for all the $x$ values, and the intensity near $E_{\rm F}$ at $x$ = 0.3 still suggests its existence; however, at $x$ = 0.2 the Fermi edge is obviously gone, which points to the topological QPT at $x_c\approx$ 0.25. This critical $x_c$ can also be inferred from the discontinuous change in the $E_{\rm F}$ intensity across $x_c\approx$ 0.25 [Fig. 3(d)]. 

   Thanks to this intensity reduction, the bulk VB becomes more visible for smaller $x$, and the peak in the EDC due to the bulk VB contribution is marked by black arrows in Fig. 3(c) (the bulk band for $x$ = 1.0 is not distinguished because the surface contribution is too strong). Note that, since those data were measured with $h\nu\sim$ 78.5 eV, the peak position in Fig. 3(c) gives the energy location of the VB maximum, and intriguingly, it is clearly dependent on $x$; in particular, the VB maximum approaches $E_{\rm F}$ as $x$ is reduced toward $x_c$, and then it moves away from $E_{\rm F}$ when $x$ passes $x_c$. This trend is summarized in Fig. 3(d) where the energy difference between the VB maximum and $E_{\rm F}$, denoted $\Delta{E}_{\rm VB}$, is plotted with error bars to take into account the finite $k_z$ and lifetime broadenings. The observed trend is consistent with the band inversion to take place at $x_c \approx$ 0.25. In view of this result, it is possible that the diminished intensity of the surface band near $x_c$ is caused by stronger interband scatterings promoted by the proximity of the bulk band.
   
   Based on the present ARPES experiments, one may draw a schematic picture of the band evolution in Pb$_{1-x}$Sn$_x$Te as shown in Fig. 4. The band inversion and the QPT takes place at $x_c\approx$ 0.25 which separates the TCI and trivial phases. The double Dirac-cone surface state is realized only in the TCI phase, in which the separation between the two Dirac cones, and hence the Dirac momentum, is dependent on $x$.
   
  Now we discuss the implications of the present experimental observations. The location of the Dirac cone in Pb$_{1-x}$Sn$_x$Te is found to be mobile in the $\Vec{k}$ space, and such a degree of freedom is obviously a unique feature of the mirror-symmetry protected TCIs, in contrast to the TIs where the Dirac point is bound to the TRIM. Intuitively, the double Dirac-cone structure in TCIs is a result of the hybridization of two Dirac cones, because, on the (001) surface, two L points (which are each responsible for a surface Dirac cone due to the band inversion at L) are projected onto the same $\bar{X}$ point; therefore, the separation of the two Dirac cones can be taken as a measure of this ``hybridization''. It is not clear at the moment what determines the strength of this ``hybridization'', but the present result seems to suggest that it has something to do with the size of the bulk band gap. Obviously, our finding is of crucial importance for establishing a concrete understanding of the physical mechanism to create the peculiar double Dirac-cone structure in TCIs.

  From the application point of view, the $x$-dependence of the Dirac momentum means that one can ``tune'' it by controlling the composition. This is particularly useful in applications requiring Fermi-surface matching with other materials. For example, to achieve an efficient spin injection, the Fermi-surface matching \cite{HaradaPRL} between the topological surface state and a ferromagnet would be crucial. The possibility to tune the location of the Dirac cone means that one may match the Dirac cone to the Fermi surface of some particular ferromagnet, thereby increasing the spin injection efficiency. 

   In summary, we performed systematic ARPES experiments on Pb$_{1-x}$Sn$_x$Te to elucidate the evolution of bulk and surface electronic states as a function of Sn concentration $x$. We found that the topological QPT between the TCI and trivial phases occurs at $x_c\approx$ 0.25 in this system and that the evolution of the energy location of the VB maximum is consistent with the band inversion taking place at this $x_c$. We also found that the separation between the two Dirac cones located across the $\bar{X}$ point becomes gradually smaller as the QPT is approached in the TCI phase. 

\begin{acknowledgments}
We thank L. Fu and S. Murakami for helpful discussions, M. Nomura, T. Shoman, K. Honma, H. Kumigashira, and K. Ono for their assistance in ARPES measurements, and T. Ueyama and K. Eto for their assistance in crystal growth. This work was supported by JSPS (NEXT Program and KAKENHI 23224010), JST-CREST, MEXT of Japan (Innovative Area ``Topological Quantum Phenomena''), AFOSR (AOARD 124038), and KEK-PF (Proposal number: 2012S2-001). M. N. acknowledges financial support from Croatian Science Foundation (Grant No. O-1025-2012). 
\end{acknowledgments}


\begin{references}
\bibitem{note} This is related to the downward band bending which somehow locates the Dirac point to be always slightly above the Fermi energy in TCI samples of Pb$_{1-x}$Sn$_x$Te, see Refs. \cite{TanakaNP} and \cite{SatoInSnTe}.
\end{references}
\end{document}